\documentclass[12 pt]{article}
\usepackage{epsfig}
\def\m{\mathrm}
\def\p{\partial}

\def\be{\begin {equation}}
\def\ee{\end {equation}}
\def\ba{\begin {eqnarray}}
\def\ea{\end {eqnarray}}
\begin{document}
\title{\textbf{Relativistic Stern-Gerlach Interaction in an RF Cavity}}
\author{M. Conte$^{[a]}$, A.U. Luccio$^{[b]}$ and M. Pusterla$^{[c]}$. \\
$^{[a]}$Dipartimento di Fisica dell'Universit\`a di Genova and
  INFN-Sezione di Genova, \\
  Via Dodecaneso 33, 16146 Genova, Italy. \\
$^{[b]}$Brookhaven National Laboratory, Upton, NY 11973, USA. \\
$^{[c]}$Dipartimento di Fisica dell'Universit\`a di Padova and
  INFN-Sezione di Padova, \\
  Via Marzolo 8, 35131 Padova, Italy.}
\date{\today}
\maketitle
\begin{abstract}
The general expression of the Stern-Gerlach force is deduced for a 
relativistic charged spin-$\frac{1}{2}$ particle which travels inside a 
time varying magnetic field. This result was obtained either by means 
of two Lorentz boosts or starting from Dirac's equation. Then, the 
utilization of this interaction for attaining the spin states separation 
is reconsidered in a new example using a new radio-frequency arrangement.
\end{abstract}
\section {The Relativistic Stern-Gerlach Force}

The time varying Stern-Gerlach, SG, interaction of a relativistic 
fermion with an e.m. wave has been proposed to separate 
beams of particles with opposite spin states corresponding to different 
energies\cite{INFN}.
We will show how spin polarized particle will exchange energy with the 
electromagnetic field of an RF resonator.

Let us denote with $(x,y,z)$ the coordinates of a particle in the 
laboratory, and with $(x',y',z')$ the coordinates in the particle rest 
frame, PRF.
In the latter the SG force that represents the 
action of an inhomogeneous magnetic field on a particle endowed 
with a magnetic moment $\vec{\mu}$ is
\be 
\textstyle
\vec f'_{SG} = \nabla'(\vec \mu^* \cdot \vec B') =
{\p \over \p x'} (\vec \mu^* \cdot \vec B') {\hat {x}} +
{\p \over \partial y'} (\vec \mu^* \cdot \vec B') {\hat {y}}  +
{\p \over \partial z'} (\vec \mu^* \cdot \vec B') {\hat {z}} 
\label {f'SG}
\ee
with
\be
\textstyle
\vec \mu = g{e\over 2m} \vec S \label {mu}. 
\ee
Here $e$ is the elementary charge with $``+"$ for protons and positrons, 
$p,e^+$, and $''=''$ for antiprotons and electrons, $\bar{p},e^-$,  
making $\vec \mu$ and $\vec S$ either parallel or antiparallel to each 
other, respectively. 
$m$ is the rest mass of the particle, $g$ the gyromagnetic ratio and 
$a$ the anomaly defined as  
\be
\textstyle
  a = {{g-2}\over 2} = 
\left\{
\begin {array} {ll} 
  1.793 ~ (g=5.586) & {\m {for}} ~ p,\bar{p}
      \\ \\
  1.160 \times 10^{-3} & {\m {for}} ~ e^\pm
\end {array}
\right..
\label {anomalies}
\ee

Notice that in Eq.(\ref{f'SG}) we have defined the magnetic moment as 
$\mu^*$ in the rest frame, rather than as $\mu'$. 
In the rest frame the quantum vector $\vec S$, or spin, has modulus 
$|\vec S|=\sqrt{s(s+1)}{\hbar}$ and its component parallel to the 
magnetic field lines can only take the following values
\be 
\textstyle
S_m=(-s,~-s+1,....,s-1,~s)\hbar, \label {S_main} 
\ee
where $\hbar$ is the reduced Planck's constant. 
Combining Eqs.(\ref{mu}) 
and (\ref{S_main}) we obtain for the magnetic moment in the PRF
\be 
\textstyle   
   \mu = |\vec \mu|=g{|e|\hbar\over 4m} =
\left\{
\begin {array} {l}
  1.41 \times 10^{-26} ~ JT^{-1}
   \\ \\
  9.28 \times 10^{-24} ~ JT^{-1}
\end {array}
\right..
\label {mumod} 
\ee 

For a particle traveling along the axis $\hat{z}$, the Lorentz 
transformations of the differential operators and of the force yield
\be 
\textstyle
\left\{
\begin{array}{lll}
{\p \over \p x'} = {\p \over \p x} &
{\p \over \p y'} = {\p \over \p y} &
{\p \over \p z'} = \gamma\,\left(
              {\p \over \p z} + {\beta\over c}
              {\p \over \p t}\right)
\\ \\
\vec f_\perp = {1\over \gamma}\,\vec f'_\perp &
\vec f_\parallel = \vec f'_\parallel &
(f_z = f'_z)
\end{array}
\label {dz}
\right..
\label{ftransf}\ee
The force ({\ref{f'SG}}) is boosted to the laboratory system as
\be 
\textstyle
\vec f_{SG} = 
{1\over \gamma} {\partial \over \partial x} (\vec \mu^* \cdot 
\vec B') 
{\hat{x}} +
{1\over \gamma} {\partial \over \partial y} (\vec \mu^* \cdot 
\vec B')
{\hat{y}} +
{\partial \over \partial z'} (\vec \mu^* \cdot \vec B') 
{\hat {z}}. 
\label {f-SG} 
\ee
Because of the Lorentz transformation of the fields\cite{Jack} 
$\vec E,\vec B$ and $\vec E',\vec B'$ 
\be
\textstyle
\left\{ 
\begin{array}{l}
\vec E' = \gamma(\vec E + c\vec \beta\times\vec B) -
{\gamma^2\over \gamma+1}\vec \beta(\vec \beta\cdot\vec E) 
\\ \\
\vec B' = \gamma\left(\vec B - {\vec \beta\over c}\times
\vec E\right) -
{\gamma^2\over \gamma+1}\vec \beta(\vec \beta\cdot\vec B)
\end{array}
\right..
\label{BR} 
\ee
the energy in the rest frame $(\vec \mu^* \cdot \vec B')$ 
becomes
\be 
\textstyle
(\vec \mu^* \cdot \vec B') = \gamma\mu^*_x
\left(B_x + {\beta\over c}E_y\right) + \gamma\mu^*_y
\left(B_y - {\beta\over c}E_x\right) + \mu^*_zB_z. 
\label{magen} 
\ee

Combining Eqs.(\ref{magen}) and (\ref{f-SG}), by virtue of
Eq.(\ref{dz}), after some algebra we can finally 
obtain the SG force components in the laboratory frame:
\be
\left\{
\begin{array}{l} 
f_x = \mu^*_x\left({\p B_x\over \p x} + {\beta\over c}
                       {\p E_y\over \p x}\right) +
          \mu^*_y\left({\p B_y\over \p x} - {\beta\over c}
                       {\p E_x\over \p x}\right) +
 {1\over \gamma}\mu^*_z{\p B_z\over \p x} 
\\  \\   
f_y = \mu^*_x\left({\p B_x\over \p y} + {\beta\over c}
                       {\p E_y\over \p y}\right) +
          \mu^*_y\left({\p B_y\over \p y} - {\beta\over c}
                       {\p E_x\over \p y}\right) +
 {1\over \gamma}\mu^*_z{\p B_z\over \p y} 
\\ \\
f_z = \mu^*_xC_{zx} + \mu^*_yC_{zy} + \mu^*_zC_{zz},    
\end{array}
\right.
\label{fSGz} 
\ee
with
\be 
\textstyle
\left\{
\begin{array}{l}
C_{zx} = \gamma^2\left[\left({\p B_x\over \p z} + 
                  {\beta\over c} {\p B_x\over \p t}\right) +
{\beta\over c}\left({\p E_y\over \p z} + {\beta\over c}
                           {\p E_y\over \p t}\right)\right] 
\\ \\
C_{zy} = \gamma^2\left[\left({\p B_y\over \p z} + 
                  {\beta\over c} {\p B_y\over \p t}\right) -
{\beta\over c}\left({\p E_x\over \p z} + {\beta\over c}
                           {\p E_x\over \p t}\right)\right] 
\\ \\ 
C_{zz} = \gamma\left({\p B_z\over \p z} + 
{\beta\over c} {\p B_z\over \p t}\right)
\end{array}
\right.. 
\label{Czz}
\ee

These results 
can also be obtained from the quantum relativistic theory
of the spin-$\frac{1}{2}$ charged particle\cite{SPIN2002}. 
Let us introduce the Dirac Hamiltonian
\be
\textstyle 
  H = e\phi + c\vec \alpha \cdot (\vec p - e\vec A) +
                                   \gamma_0 mc^2  \label {Ham1} \ee
having made use of the Dirac's matrices
\be 
\textstyle 
\vec \gamma =
       \left(\matrix {\mathcal{O}& \vec \sigma \cr -\vec \sigma & \mathcal{O} \cr}\right),
   ~~~ \gamma_0 =
       \left(\matrix {\mathcal{I} & \mathcal{O}\cr 0 & -\mathcal{I} \cr}\right),
   ~~~ \vec \alpha = \gamma_0 \vec \gamma =
       \left(\matrix {\mathcal{O} & \vec \sigma \cr \vec \sigma & \mathcal{O} \cr}\right),
                                                         \label {mat-D} \ee
where $\vec \sigma$ is a vector whose components are the Pauli's matrices
\be 
\textstyle 
    \sigma_x = \left(\matrix {0 & -i \cr i &  0 \cr}\right), ~~
    \sigma_y = \left(\matrix {1 &  0 \cr 0 & -1 \cr}\right), ~~
    \sigma_z = \left(\matrix {0 &  1 \cr 1 &  0 \cr}\right), ~~
                                                         \label {mat-P} \ee
$\mathcal{I}$ is the $2\times 2$ identity matrix, $\mathcal{O}$ the null matrix 
and having chosen the 
$y$-axis parallel to the main magnetic field.
A standard derivation leads to the non relativistic
expression of the Hamiltonian exhibiting the SG interaction with
the ``normal'' magnetic moment
\be 
\textstyle
    \tilde{H} = e\phi + \frac {1}{2m} (\vec p - e\vec A)^2 -
            \frac {e\hbar}{2m} (\vec \sigma \cdot \vec B) \label {HamP} \ee
which coincides with the Pauli equation and is valid in the PRF.

To complete the derivation we must add the contribution from the anomalous 
magnetic moment to 
the SG energy term in the previous equation, with a factor 
$1+a=\frac{g}{2}$, yielding
\be 
\textstyle
  -\frac {g}{2} \frac {e\hbar}{2m}\;\vec \sigma \cdot \vec B =
    - \vec \mu^* \cdot \vec B   ~~~~ {\mathrm {with}} ~~~~
      \vec \mu^* = g \frac {e\hbar}{4m} \vec \sigma.   \label {SGenergy}
\ee

In order to obtain the $z$-component of the SG force in the 
Laboratory frame along the direction of motion of the particle, we must 
boost the whole Pauli term of Eq.(\ref{HamP})
by using the unitary operator $U$ in the Hilbert space\cite{Hspace}, 
which expresses the Lorentz transformation
\be
\textstyle
 U^{-1} \left[g\frac{e\hbar}{4m}
       (\gamma_0\vec\sigma\cdot\vec B')\right]U =
  g\frac{e\hbar}{4m}(\gamma_0\vec\sigma\cdot\vec B')
        \left[S^{-1}(\gamma_0\sigma_x)S +
              S^{-1}(\gamma_0\sigma_y)S +
              S^{-1}(\gamma_0\sigma_z)S \right]  \label {Hilb}
\ee
that can be written in terms of the  
equivalent transformation in the $4\times4$ spinor space 
\be 
\textstyle
  S =
    \exp{\left\{\gamma_0 (\vec \gamma \cdot \hat {v}) \frac {u}{2}\right\}} =
    \cosh{\frac {u}{2}} +
    \left(\matrix {0 & \sigma_z \cr \sigma_z & 0 \cr}\right)
    \sinh{\frac {u}{2}}                                    \label {Str} \ee
with
\be 
\textstyle
    \hat {v} = \frac {\vec v}{|\vec v|}, ~~
    \cosh u = \frac {1}{\sqrt{1-\beta^2}} = \gamma = 
    {\mathrm {Lorentz~factor}},
    ~~ \sinh u =\sqrt{\gamma^2-1} ~~~ \left(\beta = \frac {v}{c}\right).
                                                        \label {defs} \ee
 
From Eqs.(\ref{Hilb}) and (\ref{Str}), due to the algebraic structure of the 
$\gamma$ and $\sigma$ matrices, we obtain in the laboratory frame the three 
components of the SG force
\be
\left\{
\begin {array}{lll}
  S^{-1} (\gamma_0 \sigma_x) S & = & \gamma_0 \sigma_x \\
  S^{-1} (\gamma_0 \sigma_y) S & = & \gamma_0 \sigma_y \\
  S^{-1} (\gamma_0 \sigma_z) S & = &
          \gamma (\gamma_0 \sigma_z) +
                   i\gamma_0\gamma_5\sqrt{\gamma^2-1} 
\end {array}
\right.,                                          
\label {evxyz}
\ee
with
\be 
\textstyle 
  \gamma_5 = \gamma_x\gamma_y\gamma_z\gamma_0 =
     i\left(\matrix{\mathcal{O} & \mathcal{I} 
	\cr \mathcal{I} & \mathcal{O}}\right).         
\label {g5} \ee

From Eqs.(\ref{evxyz}) we can deduce the expectation values of the 
SG force in the Laboratory system with a defined spin -along 
the $y$-axis in our case- via the expectation values of the Pauli matrices
and of the Pauli interaction term of the proper force
\be 
\textstyle
    f_z = \gamma_0\sigma_y\gamma^2\mu^*\left[
          \left(\frac{\partial B_y}{\partial z} + \frac{\beta}{c}
          \frac{\partial B_y}{\partial t}\right) - \frac{\beta}{c}
          \left(\frac{\partial E_x}{\partial z} + \frac{\beta}{c}
          \frac{\partial E_x}{\partial t}\right)\right].   
\label {fz4} 
\ee

In our case only the second of Eqs.(\ref{evxyz}) gives a non vanishing 
result, while both the first and third produce a null contribution to 
the force, because of the orthogonality of the 
two spin states $s=\pm \frac{1}{2}$ and the properties of the 
$\sigma$ matrices.

\section {The radio-frequency system}

\begin{figure} \centering
\epsfig{file=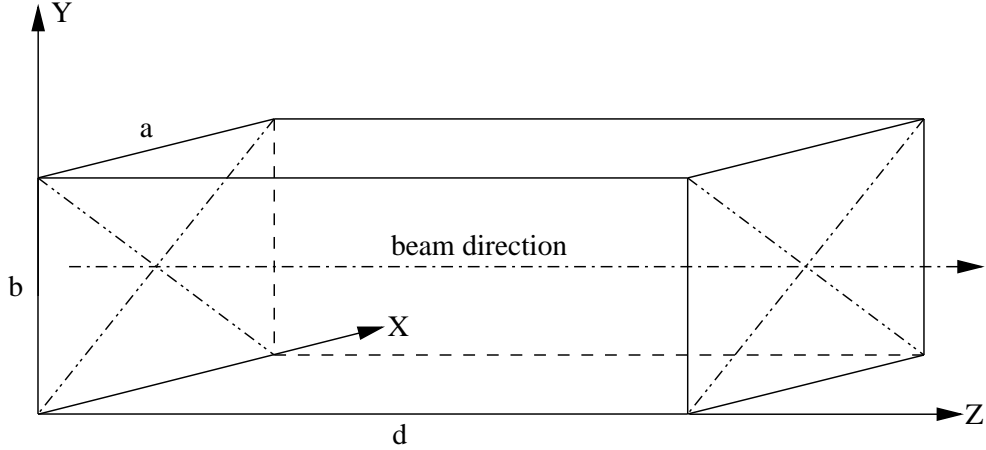,width=13cm}
\caption {Sketch of the rectangular cavity. The coordinates of the 
beam axis are $x=\frac{a}{2}$ and $y=\frac{b}{2}$.}
\label {fig:box}
\end {figure}

Let us consider the standing waves built up inside a rectangular 
radio-frequency resonator, tuned to a generic TE Mode\cite{INFN}. 
Resonator dimensions are: width $a$, height $b$ and length $d$, as shown 
in Fig.\ref{fig:box}. 
On the cavity axis, which coincides 
with the beam axis, the electric and magnetic fields 
are\cite{Ramo}
{\footnotesize
\[
\begin{array}{lcl}
E_x & = &
- B_0 \left({n\pi\over b}\right) {\omega\over K_c^2} 
 \cos\left({m\pi x\over a}\right) \sin\left({n\pi y\over b}
\right) 
 \sin\left({p\pi z\over d}\right) \sin\,\omega t     
 \\ \\
E_y & = &
  B_0 \left({m\pi\over a}\right) {\omega\over K_c^2} 
 \sin\left({m\pi x\over a}\right) \cos\left({n\pi y\over b}
\right) 
 \sin\left({p\pi z\over d}\right) \sin\,\omega t 
\\ \\ 
E_z & = & 0 ~~~~~~~~~~~~[{\m {as~typical~for~any~TE~mode}}] 
\\ \\  
B_x & = &
- {B_0\over K_c^2} \left({m\pi\over a}\right) 
\left({p\pi\over d}\right)
 \sin\left({m\pi x\over a}\right) 
\cos\left({n\pi y\over b}\right) 
 \cos\left({p\pi z\over d}\right) \cos\,\omega t 
\\ \\
B_y & = &
- {B_0\over K_c^2} \left({n\pi\over b}\right) 
\left({p\pi\over d}\right)
 \cos\left({m\pi x\over a}\right) 
\sin\left({n\pi y\over b}\right) 
 \cos\left({p\pi z\over d}\right) \cos\,\omega t 
\\ \\
B_z & = & B_0 \cos\left({m\pi x\over a}\right) 
\cos\left({n\pi y\over b}\right) 
 \sin\left({p\pi z\over d}\right) \cos\,\omega t
\end{array}
\]}
where $B_0$ is the RF peak magnetic field, 
$m$, $n$ and $p$ are integer mode indeces, and 
\be
\textstyle
K_c = \sqrt{\left({m\pi\over a}\right)^2 
+ \left({n\pi\over b}\right)^2}.           \label {kc}
\ee

The angular frequency of the e.m. wave from  the 
RF generator is 
\be
\textstyle
\omega = \omega_{\m {RF}} = {{2\pi c}\over \lambda_{\m {RF}}} = 
c\sqrt{\left({m\pi\over a}\right)^2 
+ \left({n\pi\over b}\right)^2 +
      \left({p\pi\over d}\right)^2}. 
\label{omegen} 
\ee

In contrast with an open waveguide, in a bounded cavity we can define 
a phase velocity $V_{\m {ph}}$ and a cavity wavelength $\lambda_{\m {wg}}$,
as typical of any e.m. in a refractive media, according to 
the relations 
\be
\textstyle
 \frac {V_{\m {ph}}}{c} = \beta_{\m {ph}} =
              {d\over p\pi} \sqrt{\left({m\pi\over a}\right)^2 + 
   \left({n\pi\over b}\right)^2 + \left({p\pi\over d} \right)^2}. 
\label{rphv} 
\ee
and
\be
\textstyle
\lambda_{\m {wg}} = \beta_{\m ph}\lambda_{\m {RF}}.
\label {cav-wl}
\ee

It is also
\be
  {V_{\m {ph}}} = \beta_{\m {ph}}c = 
  \beta_{\m {ph}} \frac {\lambda_{\m RF}}{\tau_{\m RF}} =
                  \frac {\lambda_{\m wg}}{\tau_{\m RF}} 
\ee

\begin{figure}
   \centering
   \epsfig{file=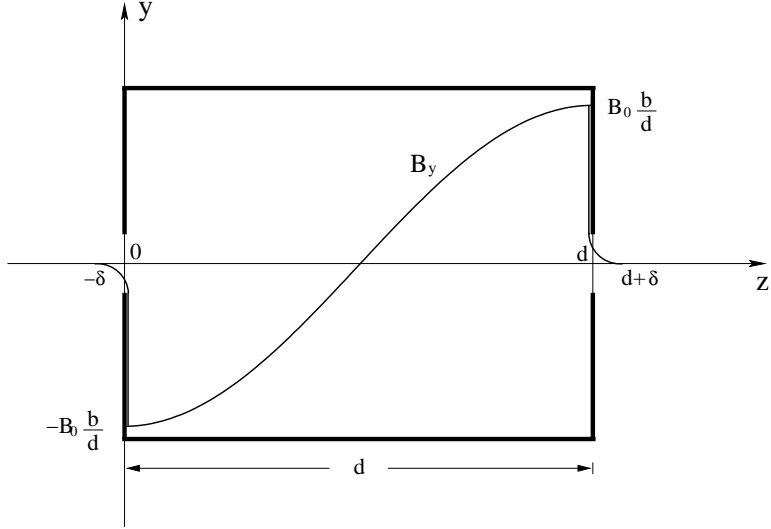,width=4in}
   \caption {Vertical component (inside the cavity) and fringes (at both 
             cavity ends) of $B_y$ for $p=1$.}
   \label{fig:edge}
\end{figure}

Notice that $\beta_{\m {ph}}$ can take any value, even larger than one, 
since it is freely dependent on the cavity geometrical parameters. 
Moreover, combining Eqs.(\ref{omegen}) and (\ref{rphv}) we obtain 
\be
\textstyle
  d = \frac {1}{2}p\beta_{\m {ph}}\lambda_{\m {RF}} = 
      \frac {1}{2}p\lambda_{\m {wg}} 
\label {cav-length} 
\ee
\par\noindent
which describes the connection between the cavity length 
$d$ and the wavelengths, as shown in Fig.\ref{fig:edge}. 
For simplicity, let's choose the transverse 
electric mode ${\m {TE}}_{01p}$, so Eqs.(\ref{omegen}) and (\ref{rphv}) 
reduce respectively to
\be
\textstyle
  \omega = \omega_{\m {RF}} = c\sqrt{\left(\frac{\pi}{b}\right)^2 +
                 \left(\frac{p\pi}{d}\right)^2 } = 
                 \beta_{\m {ph}}c\frac{\pi}{d}
          ~~~~ {\m {and}} ~~~~
  \beta_{\m {ph}} = \sqrt{1+\left(\frac{pd}{b}\right)^2}
                                                   \label{ombetap} 
\ee
or, setting the mode index $p=1$,
\be
\textstyle
  \omega = \omega_{\m {RF}} = c\sqrt{\left(\frac{\pi}{b}\right)^2 +
                 \left(\frac{\pi}{d}\right)^2 } = 
                 \beta_{\m {ph}}c\frac{\pi}{d}
          ~~~~ {\m {and}} ~~~~
  \beta_{\m {ph}} = \sqrt{1+\left(\frac{d}{b}\right)^2},
                                                   \label{ombeta1} 
\ee
which are the quantities pertaining to the preferred ${\m {TE}}_{011}$ 
mode whose non zero field components on the cavity axis are 
\be
\textstyle
\left\{ \begin{array}{l}
  B_y(z,t) = - B_0 \frac{b}{d}\cos\left(\frac{\pi z}{d}
\right)\cos\omega t 
\\ \\
  E_x(z,t) = -\omega B_0 
\frac{b}{\pi}\sin\left(\frac{\pi z}{d}\right)
                  \sin\omega t
\end{array} \right..                              
\label {ByEx}  
\ee

It is important to emphasize that in all the field
components met so far there is a clear separation between spatial and 
temporal contributions, as typical of standing waves. Besides, the 
boundary conditions of the electric and magnetic fields of the e.m. 
dictate the shape of the spatial component 
which, in turn, oscillates in time with the frequency $\omega_{RF}$.
Then, at the cavity entrance and exit the field components ~(\ref{ByEx}) 
become on axis
\be
\textstyle
{\m {Entrance}} \Longrightarrow 
\left\{ \begin{array}{l}
B_y(0,t) = -B_0\frac{b}{d}\cos\omega t 
\\ \\
E_x(0,t) = 0
\end{array} \right..
\label{edge-in}
\ee
and 
\be
\textstyle
{\m {Exit}} \Longrightarrow 
\left\{ \begin{array}{l}
B_y(d,t) = -B_0\frac{b}{d}\cos\pi\cos\omega t
                      = B_0\frac{b}{d}\cos\omega t
\\ \\
E_x(d,t) = -\omega\frac{b}{\pi}\sin\pi\sin\omega t = 0
\end{array} \right..
\label{edge-out} 
\ee
where $t$ is a generic time. The null values of $E_x$ at the cavity ends 
confirm a typical pattern of the transverse electric mode.

\section{Stern-Gerlach interaction with the cavity field}

From Eq.(\ref{fz4}), after some algebra, we obtain tha a charged fermion which crosses 
a radio-frequency resonator, tuned
on the {TE}$_{011}$ mode, acquires (or loses) an energy amount 
when interacts with the field component in the ``body'' of the cavity 
shown in Fig.~\ref{fig:edge}\cite{INFN} 
\be
  (\delta U)_{\m {X-ing}} = \int_0^d\,f_z dz = \int_0^d\,\mu^*C_{zy}dz =
   \beta^2\gamma^2 B_0\mu^* \frac{b}{d}
   \frac {\beta_{\m {ph}}^2(2-\beta^2)-1}{\beta_{\m {ph}}^2-\beta^2 }
   \left(1+\cos\frac{\beta_{\m {ph}}}{\beta}\pi\right)   \label {dU-X}
\ee
still assuming that the spin is not precessing. 

However, since the cavity cannot be completely enclosed but must have 
apertures at both ends to allow the particle bean to pass through and 
consequently will have fringe fields, in order to calculate the full 
SG interaction it is necessary to deal with the interaction with these 
fields. This is discussed right below.
\subsection{Fringe fields}

\begin{figure} 
\centering
\epsfig{file=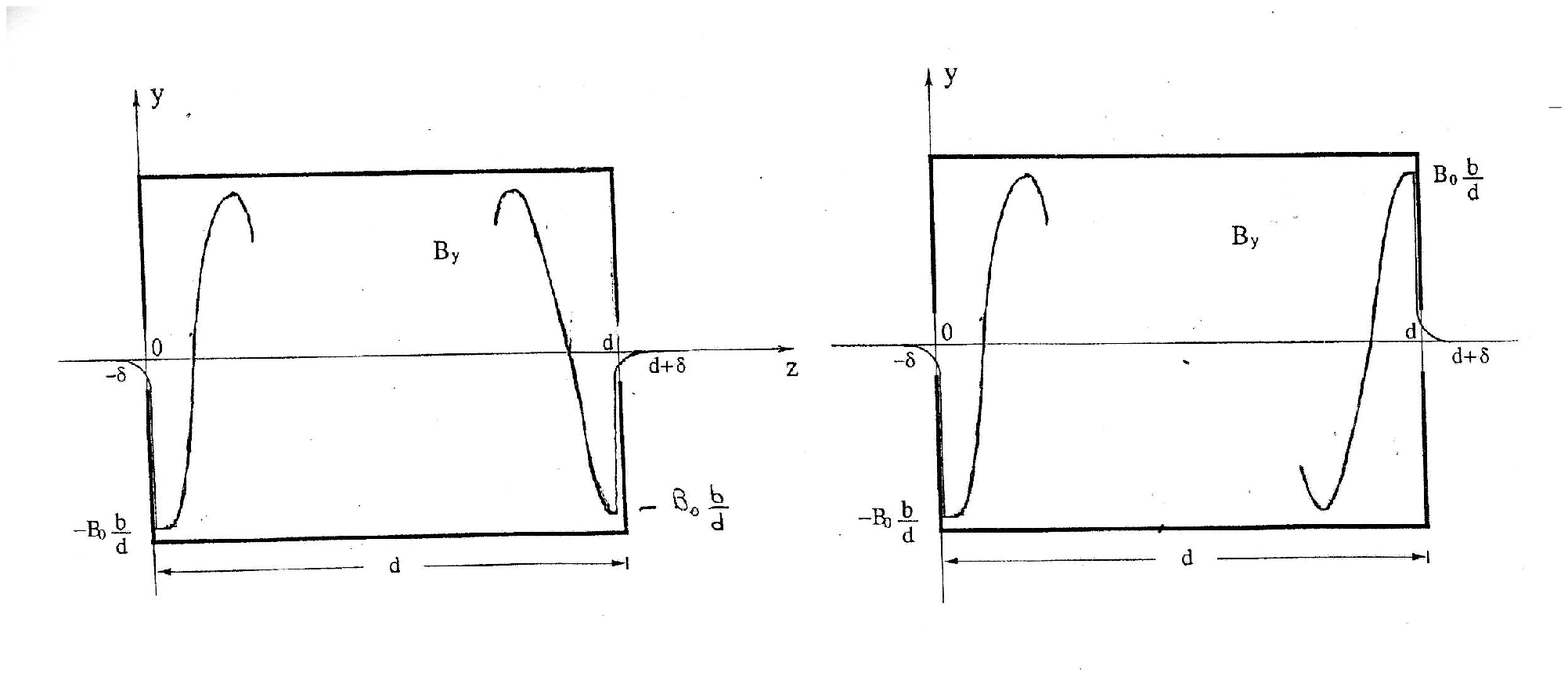,width=13cm}
\caption{Edge fields at both ends of a single cavity for $p$
equal to either an even number (left) or an odd number (right).}
\label{fig:edges}
\end{figure}

In order to fulfill the boundary conditions (\ref{edge-in}) and 
(\ref{edge-out}), a cavity tuned in its ${\m {TE}}_{011}$ mode must be 
exactly filled by either an even or an odd number of cavity dependent 
half wave-lengths, Eq.(\ref{cav-wl}), as illustrated in 
Figs.~\ref{fig:edge} and \ref{fig:edges}.

Consider now a bunch of particles crossing the cavity in synchronism 
with the RF field. 
This requires  that the bunch centre of mass that enters the cavity at 
the instant $t=0$ and would leave the cavity at $t=\tau_{\m {RF}}$, 
at magnetic field values, respectively
\be
\textstyle
  B_y(0,0) = -B_0\frac{b}{d}  ~~~~ {\m {and}} ~~~~
  B_y(d,\tau_{\m {RF}}) = B_0\frac{b}{d}           \label {in-out}
\ee

The field values at both ends fade rapidly to zero over a 
small distance $|\delta|$ just outside the cavity (see Figures.) 
We may consider these fringe fields as small-valued  
functions in the $(y,z)$-plane, since the time $\delta t$ 
necessary for a particle to proceed through this distances can be very 
small in comparison with $\tau_{\m {RF}}$, depending of course by the 
size of the beam channel, or 
\be
\textstyle
\left\{
\begin{array}{llllll} 
\left[B_y\right]_{\m {in}} &=& -B_0\frac{b}{d}\;g(z) 
&{\m {with}} &
g(-\delta)=0, & g(0)=1
\\ \\
\left[B_y\right]_{\m {out}} &=& B_0\frac{b}{d}\;h(z) 
&{\m {with}} &
h(d)=1, & h(d+\delta)=0
\end{array}
\right..
\label{edge-flds} 
\ee

Under these conditions, a relativistic fermion with its spin 
directed along the $y$-axis and traversing  the cavity will experience 
a SG force parallel to the $z$-axis (direction of motion),
see Eq.(\ref{fSGz}) 
\be 
\textstyle
f_z = \mu^* C_{zy} 
\label{fz} 
\ee
where $C_{zy}$ is given by the second of the set of Eqs.(\ref{Czz}). 
For the moment we assume that the spin will conserve its orientation 
during traversal

The electric field $E_x$ and its
derivatives in this equation are almost constantly zero, 
because of the boundary conditions on the walls of the cavity and at 
the extreme points $z=-\delta$ and $z=d+\delta$. 
Furthermore, the function$\left(\frac{\partial B_y}{\partial t}\right)$ 
is almost zero along the fringe
segments because of its proportionality to $\sin\omega t$,  
with $t$ equal to the $\delta t$ mentioned before. 
Consequently we have
\be
\textstyle
C_{zy} \simeq \gamma^2 \frac{\p B_y}{\p z},
\label {redCzy} \ee
and for the entire fringe field 
\be
\textstyle
\left\{
\begin{array}{lll}
\left[f_z\right]_{\m {in}} &=& -B_0\mu^*\frac{b}{d}\gamma^2 
                           \left(\frac{dg(z)}{dz}\right)
\\ \\
\left[f_z\right]_{\m {out}} &=& B_0\mu^*\frac{b}{d}\gamma^2 
                           \left(\frac{dh(z)}{dz}\right)
\end{array}
\right..
\label{edge-forces} 
\ee

Making use of eqs.~(\ref{redCzy}) and (\ref{edge-forces}),  
the energy 
increments $[\delta U]_{\m in}$ and $[\delta U]_{\m out}$ 
related to the 
fringe fields are easily evaluated since the integrals 
$\int_{-\delta}^{0}f_z dz$ and $\int_{d}^{d+\delta}f_z dz$ 
only depend 
upon the extreme points~(\ref{edge-flds}) and do not depend 
on the curve that connects them. 
In fact $f_z dz$ becomes an exact differential. 
Then we obtain for the energy exchange at both edges 
\be
\textstyle
(\delta U)_{\m {in}} = (\delta U)_{\m {out}} 
= - B_0\mu^*\frac{b}{d}\gamma^2.
\label{cav-en} 
\ee

The total energy exchange at the edges is therefore
\be
\textstyle
(\delta U)_{\m {ff}} =
(\delta U)_{\m {in}} + (\delta U)_{\m {out}} 
= - 2B_0\mu^*\frac{b}{d}\gamma^2.
\label {ff-en}
\ee
\subsection{Full energy interaction}

By adding the fringe contributions~(\ref{ff-en}) to the cavity body 
crossing contribution~(\ref{dU-X}) seen before, obtain
\be
  (\delta U)_{\m {tot}} = (\delta U)_{\m {ff}} + (\delta U)_{\m X} =
  -\gamma^2 B_0\mu^*\frac{b}{d} f(\beta_{\m {ph}},\beta)
                                               \label {dU-tot} 
\ee
with
\be
\textstyle
  f(\beta_{\m {ph}},\beta) = \left[2-\beta^2
  \frac {\beta_{\m {ph}}^2(2-\beta^2)-1}{\beta_{\m {ph}}^2-\beta^2 }
  \left(1+\cos\frac{\beta_{\m {ph}}}{\beta}\pi\right)\right].  
                         \label {beta-beta}
\ee

For ultra relativistic particles ($\beta\simeq 1$) Eq. (\ref{dU-tot}) reduces to
\be
  (\delta U)_{\m {tot}} \simeq  -\gamma^2 B_0\mu^*\frac{b}{d}
   (1-\cos\beta_{\m {ph}}\pi). 
                                            \label {dU-UR1}  \ee

This last result deserves a few comments. 
In fact, if we set 
\be 
\textstyle
  \beta_{\m {ph}}=2 ~~~ \Longrightarrow ~~~
d = \frac {1}{2}\lambda_{\m {wg}} = \lambda_{\m {RF}}
  \label {b2} \ee
the total energy contribution (\ref{dU-UR1}) vanishes, implying a full 
cancellation of the effect. 

On the other hand if we set
\be 
\textstyle
  \beta_{\m {ph}}=3 ~~~ \Longrightarrow ~~~
       d = \frac {1}{2}\lambda_{\m {wg}} = \frac {3}{2}\lambda_{\m {RF}}
  \label {b3} \ee
the total energy contribution (\ref{dU-UR1}) becomes
\be
\textstyle
  (\delta U)_{\m {tot}} \simeq  -2\gamma^2 B_0\mu^*\frac{b}{d}
                                             \label {dU-UR2} \ee
as deduced from Eq.(\ref{cav-length}). 
In Table I we gather values calculated from Eq.(\ref{beta-beta})  
for non-relativistic and ultra-relativistic particles for, either 
$\beta_{\m {ph}}=2$ or $\beta_{\m {ph}}=3$ at two proton energies.  
Each $\beta_{\m {ph}}$ is accompanied by the corresponding ratio 
cavity-length over cavity-height.
\begin {center}
{\bf Table I:} $f(\beta_{\m {ph}},\beta)$
\par\vspace {0.5 cm}\noindent
\begin{tabular}{|c|c|c|}
\hline
 $\beta_{\m {ph}}\Rightarrow\frac{d}{b}$  & Low Energy  &   High Energy  \\
    &  (e.g. $W_{\m {kin}}$ = 5 MeV) & (e.g. $W_{\m {kin}}$ = 30 GeV) \\
\hline
\hline 
     2 $\Rightarrow$ 1.732   &    2.01   &    0      \\
\hline     
     3 $\Rightarrow$  2.828  &    2.02   &    2      \\
\hline
\end{tabular}
\end {center}

\begin{figure} 
\centering
\epsfig {figure=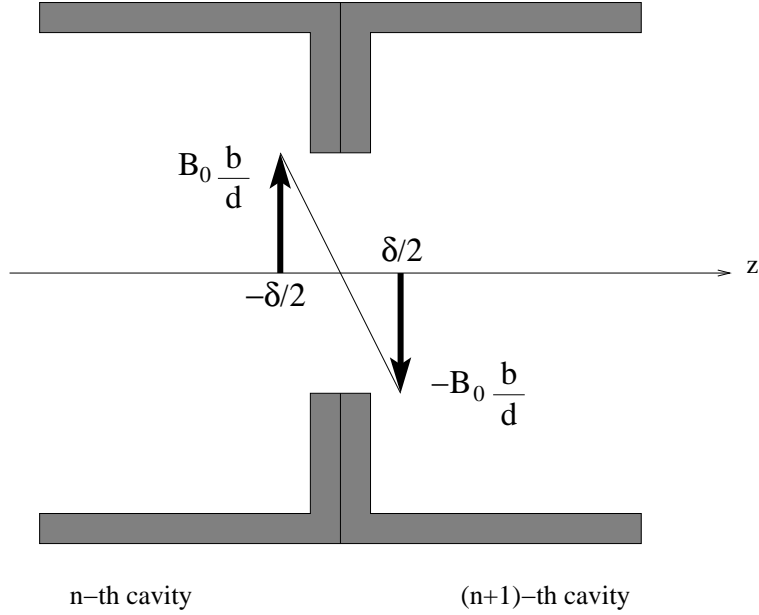,width=10cm}
\caption{Magnetic field gradient between two contiguous cavities}
\label{fig:cavX}
\end{figure}

Furthermore, if we consider two contiguous cavities, there will be a 
gradient between the positive $B_y$ at the end of the first cavity and 
a negative $B_y$ at the beginning of the second cavity, as shown in 
Fig.~\ref{fig:cavX}. 
In this case we may consider the magnetic field at the interface as 
linearly dependent on $z$, that is
\be
\textstyle
[B_y(z)]_{\m {X-ing}} = -2B_0\frac{b}{d\delta}z.
\label{gradX} 
\ee

Reiterating what done before, obtain
{\footnotesize
\be
\begin{array}{rcl}
\frac{\p}{\p z}\left[B_y(z)\right]_{\m {X-ing}} & 
= & -2B_0\frac{b}{d\delta} 
\\ \\
f_z & = & -2B_0\mu^*\frac{b}{d\delta}\gamma^2 
\\ \\
(\delta U)_{\m {cav2cav}} = (\delta U)_{\m {cc}}   & 
= & \int_{-\frac{\delta}{2}}^{\frac{\delta}{2}}f_z dz =
-2B_0\mu^*\frac{b}{d}\frac{1}{\delta}
\left[\frac{\delta}{2}
-\left(-\frac{\delta}{2}\right)\right]\gamma^2 =
-2B_0\mu^*\frac{b}{d}\gamma^2 
\end{array}
\ee }
\par\noindent
which means that, for $N$ cavities, we shall have as 
final result for ultra relativistic particles 
{\footnotesize
\be
\textstyle
(\delta U)_{\m {tot}} = N(\delta U)_{\m {X-ing}} 
      - (N-1)(\delta U)_{\m {cc}} 
      - (\delta U)_{\m {ff}} =
\left\{
\begin {array} {ll}
  0 & {\m {for}} ~ \beta_{ph} = 2
   \\ \\
 \frac {2N}{2.83} B_0\mu^*\gamma^2 & {\m {for}} ~ \beta_{ph} = 3
\end {array}
\right..
\label{full-end} 
\ee }

Conversely, if $\beta_{\m {ph}}$ is even, particles with their spin 
pointing always in the same direction cannot exchange energy with the
standing wave of a TE resonator. 
A spin rotator\cite{PAC07} can align the particle magnetic moments 
either parallel or anti-parallel to the 
directions of the magnetic field gradients, thus allowing the desired
energy interaction. 
This situation would be similar to what happens 
in a multi-stage tandem van de Graaff, where the ions are repetitively 
accelerated by the same electrostatic field, becoming alternatively 
negative, via an addition of electrons, or positive, via electron 
stripping.

Unfortunately, the field integral 
($B_{\m M}d = \beta\pi\frac{mc}{ae}=\beta\;5.46$ Tm, for $p,\bar{p}$) 
for attaining a spin rotation is so large that this solution is 
unpractical. 
Instead, the example of $\beta_{\m {ph}}$ equal to an odd number 
seems much more suitable since does not require cumbersome magnets, 
but only longer cavities (compare Eqs. (\ref{b2}) 
and (\ref{b3})). In fact, the magnetic moments are (de)accelerated by the
field tails at the cavity ends, while don't change their energy when 
crossing the cavities. This situation resembles the Wideroe linac where 
the charged particles are accelerated by the electric fields between two
contiguous drift tubes, but don't change their energy while crossing 
the tubes themselves.
\section {Concluding remarks}

On the basis of the previous estimates, we feel ready to propose the 
time varying SG interaction as a method for attaining a spin 
state separation of an unpolarized beam of, say (anti)protons, since 
the energy of particles with opposite spin orientations will differ and 
beams in the two states can be separated. 
In a first stage of the study of a sensible practical design, we intend 
to proceed with numerical simulations. 
As a first step, we intend to verify the correctness of 
Eqs.(\ref{dU-tot}) and (\ref{beta-beta}) setting once 
$\beta_{\m {ph}}=2$ and then $\beta_{\m {ph}}=3$, in a cavity where 
the field line pattern can be realistically controlled. 

Beyond the verification of the present theory, there is also the aim of 
studying the effects generated by the spin precession inside the 
cavity, that we did not yet address in this note. 

Next, we shall consider a spin splitter scheme based on the lattice of 
an existing or planned (anti)proton ring endowed with an array of 
splitting cavities. 
The principal aim of the latter implementations is to check the mixing 
effect\cite{CMP}\cite{Marco} of the longitudinal phase-plane filamentation, 
i.e. the actual foe which could frustrate the entire spin splitting process.

\section {Acknowledgments}

First, we want to thank Waldo MacKay, who has participated on so many 
discussions on the whole idea but who was regrettably prevented by 
numerous commitments from participate to the editing of the present note.
We thank Renzo Parodi for his help for us to better understand the 
subtleties of the standing waves building up. 
Thanks are also due to Chris Tschalaer for fruitful discussions on the role of the fringe fields.


\begin{thebibliography}{}

\bibitem {INFN}
M. Conte, M. Ferro, G. Gemme, W.W. MacKay, R. Parodi, M. Pusterla:
The Stern-Gerlach Interaction Between a Traveling Particle and a
Time Varying Magnetic Field, INFN/TC-00/03, 22 Marzo 2000.
(http:xxx.lanl.gov/listphysics/0003, preprint 0003069)
%
\bibitem {SPIN2002}
P. Cameron, M. Conte, A. Luccio, W.W. MacKay, M. Palazzi and 
M. Pusterla: 
The Relativistic Stern-Gerlach Interaction and Quantum Mechanics 
Implications,
Proceedings of the SPIN2002 Symposium, 9-14 September 2002, 
Brookhaven, 
Eds. Y.I.
Makdisi, A.U. Luccio and W.W. MacKay, 
AIP Conference Proceedings 675 (2003) p. 786.
%
\bibitem {Jack}
J.D.Jackson, {\it Classical Elecrodynamics}, John Wiley \& Sons Inc., 
New York 1975
%
\bibitem {Hspace}
R.P. Feynman, {\it Quantum Electrodynamics}, W.A. Benjamin Inc., New York 1961. 
%
\bibitem {Ramo}
S. Ramo, J.R. Whinnery and T. Van Duzer, {\it Fields and Waves in
Communication Electronics}, John Wiley and \& Sons, New York, 1965.
%
\bibitem {PAC07}
M.Conte,A.U.Luccio,W.W.MacKay and M.Pusterla
{\it Stern Gerlach Force on a Precessing Magnetic Moment}
Proc. PAC07, Albuquerque, NM (2007), p.3729
%
\bibitem {CMP}
M. Conte, W.W. MacKay and R. Parodi: An Overview of the Longitudinal 
Stern-Gerlach Effect, BNL-52541, UC-414, November 17 1997.
%
\bibitem {Marco}
M. Palazzi: Ph.D Thesis, Genoa University, June 6 2003.

\end {thebibliography}
\end{document}